\documentclass[prc,twocolumn,showpacs,superscriptaddress]{revtex4-2}
\usepackage{dsfont}
\usepackage{graphicx,amsmath}
\usepackage{natbib}
\usepackage{xcolor}
\def\be{\begin{eqnarray}}
\def\ee{\end{eqnarray}}
\def\bc{\begin{center}}
\def\ec{\end{center}}

\newcommand{\lsim}{\stackrel{\scriptstyle <}{\phantom{}_{\sim}}}
\newcommand{\gsim}{\stackrel{\scriptstyle >}{\phantom{}_{\sim}}}

\begin{document}
\title{The NICER data and a $\sigma$-field dependent stiffness of the hadronic equation of state}
\author{E. E. Kolomeitsev}
\affiliation{Bogoliubov Laboratory of Theoretical Physics, Joint Institute for Nuclear Research, RU-141980 Dubna, Russia}
\affiliation{Matej Bel  University, SK-97401 Banska Bystrica, Slovakia}
\author{D. N. Voskresensky}
\affiliation{Bogoliubov Laboratory of Theoretical Physics, Joint Institute for Nuclear Research, RU-141980 Dubna, Russia}
\affiliation{National Research Nuclear  University ``MEPhI'', Kashirskoe  Av. 31, RU-115409 Moscow, Russia}
\begin{abstract}
Analyses for the NICER data indicate that there is no significant variation of the compact star radii within the mass range of 1.4 to 2.0 solar masses.  Yamamoto et al. [Phys.\ Rev.\ C \textbf{108}, 035811 (2023)] concluded recently that ``this feature cannot be reproduced by the hadronic matter due to the softening of the equation of state (EoS) by hyperon mixing, suggesting the possible existence of quark phases in neutron-star interiors.'' Using a collection of 162 purely nucleonic, hyperonic, and quarkish EoSs from CompOSE database and some other works, we verify that hyperons indeed lead to a significant difference in radii of stars of 1.4 and 2.0 solar masses, which diminishes in the presence of quarks. We compare the shapes of the mass-radius curves and show that hyperons and quarks in the neutron star cores prefer a particular curve shape with backbending. It is argued that the shape {is controlled by the density dependence} of the nuclear symmetry energy. We draw attention to the existence of a class of purely hadronic relativistic mean-field EoSs with scalar-field dependent hadron masses and coupling constants that satisfy the known constraints on the EoSs including the analyses of the new NICER data and the above requirement of no significant variation of the neutron star radii.
\end{abstract}
\keywords{neutron stars, CompOSE, equation of state, NICER, mass-radius curve}
 \maketitle

\section{Introduction}

The advent of the multi-messenger astronomy enables studies of neutron stars (NSs) through all the available tracers:  cosmic rays, neutrinos, electromagnetic  and gravitational waves. New data provide more stringent constraints on the main NS parameters, which are used to get insight into  hadron/nucleon interactions at supra-saturation densities~\cite{Lattimer-ARNPS21}. Thanks to dedicated campaigns of the radio pulsar timing measurements several heavy NSs with masses greater than two solar masses were identified. The recently launched X-ray timing telescope, the Neutron Star Interior Composition Explorer (NICER), delivered several first joint measurements of star masses and radii. Based on the attempts to describe the new data, works have appeared with the conclusion that purely hadronic equation of state (EoS) cannot fully account for them~\cite{Baym-TKPST-RPP81,Yamamoto:2023osc}, and  that a hydrid EoS involving sub-hadronic degrees of freedom (quarks, diquarks) is needed.

The purpose of this paper is, first, to restate, which of the NS properties may appear problematic for the description with the purely hadronic EoSs, and, second, to point out a class of hadronic models that accommodate the new data.

\section{Masses of neutron stars}

For quite some time there has been a consensus that most of NSs have masses nearby
$1.4\,M_{\odot}$~\cite{Thorsett-ApJ512-NSMass14}, being produced in supernova explosions with masses close to the NS maximum mass $M_{\rm max}\simeq 1.5\,M_\odot$~\cite{Brown-Bethe-ApJ423}. Most of the existed hadronic EoSs could describe NSs with such masses. Also such a very narrow NS mass distribution offered a convenient way to explain the NS cooling data within the minimal cooling plus direct Urca (DU) scenario~\cite{Page-ApJSup155-minvcool,Yakovlev-GKLP-ASP33} without including in-medium effects. In contrast, a nuclear medium cooling scenario was developed in Refs.~\cite{Voskresensky:1986af,Schaab:1996gd,Voskresensky:2001fd,Blaschke:2004vq}. It relied on the assumption that NSs with measured surface temperatures (first data fixed only upper limits on surface temperatures) have very different masses and that NS neutrino emissivity depends strongly on the density--NS mass, since the in-medium pion exchange significantly affects the two-nucleon reaction rates, whereas the DU reaction is not allowed~\cite{Voskresensky:1984zzn,Voskresensky:1986af}.  This approach has been supported by the discovery of a light pulsar with mass $1.25\,M_\odot$ in the double pulsar system J0737-3039~\cite{Lyne-Science303-Double-NS} and by the growing evidence for the existence of NSs with masses greater than $1.5\,M_\odot$~\cite{Nice-ApJ634}. It is now well established that the NS masses vary over a wide interval~\cite{Lattimer-ARNPS62}. The so-far lightest NS \footnote{Recent estimate for the CCO XMMU~J173203.3-344518 mass, $0.77^{+0.20}_{-0.17} M_{\odot}$, reported in Ref.~\cite{Doroshenko-light-HESS} will be discussed below.} with the well-measured mass of $1.174(4)\,M_{\odot}$ is the pulsar PSR~J0453+1559~\cite{Martinez2015}. The masses of the heaviest pulsars are mainly derived from analyses of Shapiro delay measurements of pulsar binaries. The first well-measured masses were $1.908(16)\,M_{\odot}$ of PSR J1614-2230~\cite{Demorest2010,Fonseca2016,Arzoumanian2018} and $2.01(4)\,M_{\odot}$ for PSR J0348+0432~\cite{Antoniadis2013}. The current highest precisely measured mass is $2.08(7)\,M_{\odot}$ for PSR J0740+6620~\cite{Cromartie2020,Fonseca2021}. Additional information is obtained from the photometry of binary systems of millisecond pulsars in tight $<1$ day orbits, with the companion heated and evaporated by the pulsar spindown power -- the so-called spiders: black widows with substellar companions and redbacks with low-mass star companions. Among these objects there is the fastest rotating pulsar PSR~J0952-0607~\cite{Romani2022} whose mass is found to be $2.35(17)\, M_{\odot}$. A joint analysis including other spider pulsars in~\cite{Kandel_2023} leads to the conclusion that the minimum value for the maximum NS mass is $M_{\rm max} > 2.19\,M_\odot$ with $1\sigma$ confidence. It is to be noticed that all spider NSs are fast-rotating millisecond pulsars so one should include corrections for a possible increase of the star mass due to rotation, which is estimated in~\cite{Brandes:2023hma} as $3\%$. Therefore, the lower limit on the maximum NS mass should be lowered and, consequently, we would have $M_{\rm max}>2.1 M_{\odot}$. There is also a constraint on $M_{\rm max}$ from above. The authors of Ref.~\cite{Rezzolla-ApJL852-Mmax} combining gravitational wave observations of merging systems of binary NSs and quasi-universal relations concluded that for a non-rotating NS the maximum mass should satisfy the constraint, $M_{\rm max}<2.33\,M_\odot$. If so, the value of $M_{\rm max}$ can be considered as a very-well constrained quantity, $2.1 M_{\odot}<M_{\rm max}<2.33\,M_\odot$.

\section{Hyperon and $\Delta$ puzzles}

Many purely nucleonic EoSs can satisfy the constraint $M_{\rm max}>2.1\,M_{\odot}$, see Ref.~\cite{Klahn-PRC74} and review of Skyrme models~\cite{Stone-Miller-PRC68-Skyrme}. In the relativistic mean-field (RMF) modified Walecka models this inequality can be easily satisfied by choosing a sufficiently small effective nucleon mass at the saturation density as an input parameter, see Fig.~6 in Ref.~\cite{KV-NPA759-KVOR}.
However, allowing for the existence of strange particles in the model and the population of the corresponding Fermi seas, one realizes
\cite{Schaffner-NPA804-hyppuzzle,Djapo-PRC81-hyperon} that employing an empirically motivated two-body hyperon–-nucleon potential, hyperons appear in NS matter already at baryon densities $\gsim (2-3)\,n_0$, where $n_0\simeq 0.16$\,fm$^{-3}$ is the nuclear saturation density. The coupling constants of hyperons with vector mesons were interrelated by SU(6) symmetry relations, cf.~\cite{vanDalen-Sedrakian-PLB734}. As a result, the maximum masses of NSs with hyperons fall below not only $2\,M_{\odot}$ but also below $1.4\,M_{\odot}$. This was called in the literature as
the ``hyperon puzzle'', which can be avoided by artificially preventing the appearance of hyperons or by including the hyperon--nucleon and/or hyperon--hyperon density-dependent repulsion, e.g., due to three-body forces~\cite{Petschauer:2020urh}, see the discussion in Ref.~\cite{Weise:2023cua}. In the framework of RMF models one can include the hyperon-hyperon repulsion mediated by a $\phi$-meson mean field and/or use a different choice of hyperon-meson coupling constants beyond the quark counting within the SU(6) symmetry, see, e.g., Ref.~\cite{Weissenborn:2011ut}, to increase the NS mass. The similar ``$\Delta$ puzzle'' with the occupation of $\Delta$ isobar Fermi seas was identified in Ref.~\cite{Drago-PRC90-Delta}.

Another aspect of the hyperon puzzle is that the presence of hyperons in the NS interiors allows efficient DU reactions on hyperons (HDU), e.g. $\Lambda \to p+e+\bar{\nu}$ leading to very fast cooling of NSs with masses $M>M_{\rm HDU}$, where $M_{\rm HDU}$ is the NS mass, at which the first hyperons appear in the star center. This second part of the problem is solved within the nuclear medium cooling scenario in~\cite{Grigorian:2017xqd,Grigorian:2018bvg}.

\section{Radii of neutron stars}

In the early 2000s, experimental data on the NS radii began to appear: from analyses of  quasi-periodic oscillations in the low-mass X-ray binary system  4U~0614+09~\cite{van-Straaten-ApJ540}, the thermal emission of the bright isolated NS RX J1856.5-3754~\cite{Trumper-NPB132}, thermonuclear X-ray bursts from NSs in low-mass X-ray binaries \cite{Oezel-ApJ820,Suleimanov-MNRAS466}, and pulse-phase-resolved X-ray spectroscopy~\cite{Bogdanov_2013,Hambaryan_2014}. See also the Bayesian analysis of combined data in~\cite{Lattimer_2014}. In most of these works, the masses of the studied objects were poorly constrained and only broad regions on the mass-radius plane (in some cases not even overlapping) were marked as allowed. This situation began to change with the launch of the NICER observatory. In its first measurement campaigns NICER studied the millisecond pulsar PSR~J0030+0451 whose mass was found in two independent analyses to be $1.34^{+0.15}_{-0.16}\,M_{\odot}$~\cite{Riley_2019} and $1.44_{-0.14}^{+0.15}\,M_\odot$~\cite{Miller_2019} and the inferred radius was determined to be $12.71^{+1.14}_{-1.19}$\,km in~\cite{Riley_2019} and $13.02_{-1.06}^{+1.24}$\,km in~\cite{Miller_2019}. NICER then turned to one of the heaviest NSs, object PSR J0740+6620. The radius was found to be $13.7^{+2.6}_{-1.5}$\,km in~\cite{Miller_2021} and $12.39^{+1.30}_{-0.98}$\,km in \cite{Riley_2021}.
{Applying the two-star radius measurements with the tidal deformability constraints to three different frameworks for EoS, Ref.~\cite{Miller_2021} provided the following 68\% credible intervals of the radius estimates
\begin{align}
R_{1.4\,M_\odot}=12.45(65)\,{\rm km} \,,\,\, R_{2.0\,M_\odot}=12.35(75)\,{\rm km}\,.
\label{Miller}
\end{align}
}
The NICER data~\cite{Riley_2019,Miller_2019,Miller_2021,Riley_2021} have been incorporated in Ref.~\cite{Legred_2021} into the joint analysis of the NS EoS, using a nonparametric EoS model based on Gaussian processes and combining information from other X-ray, radio and gravitational wave observations of NSs. The results are
\begin{align}
&R_{1.4\,M_\odot}= 12.56^{+1.00}_{-1.07}\,{\rm km}\,\mbox{\cite{Miller_2019}}\,\mbox{and}\, 12.34^{+1.01}_{-1.25}\,{\rm km}\,\mbox{\cite{Riley_2019}}\,,
\label{Legred-14}\\
&R_{2.0\,M_\odot}= 12.41^{+1.00}_{-1.10}\,{\rm km}\,\mbox{\cite{Miller_2021}}\,\mbox{and}\,  12.09^{+1.07}_{-1.17}\,{\rm km}\,\mbox{\cite{Riley_2021}}.
\label{Legred-20}
\end{align}
These analyses show that despite significant statistical uncertainties, the derived NS radii are consistent with being equal over a wide mass range, with a radius difference of
\begin{align}
&\Delta R_{(1.4-2.0)M_{\odot}} \equiv R_{1.4M_\odot} - R_{2.0M_\odot}
\nonumber\\
&\qquad=
\left\{ \begin{array}{cc}
0.12_{-0.83}^{+0.85}\,{\rm km} & \mbox{(Miller et al.~\cite{Miller_2019,Miller_2021})}\\
0.20_{-0.82}^{+0.8}\,{\rm km} & \mbox{(Riley at al.~\cite{Riley_2019,Riley_2021})}
\end{array}
\right. ,
\label{Legred-DR}
\end{align}
see Table~2 in~\cite{Legred_2021}. The results of the combined analyses collected in Table~4 in Ref.~\cite{Miller_2021} assume that $-0.48\,{\rm km}\lsim \Delta R_{(1.4-2.0)M_{\odot}}\lsim 0.35\,{\rm km}$. The lower limit $\Delta R_{(1.4-2.0)M_{\odot}} \simeq -0.68\,{\rm km}$ follows from the direct NICER measurement by Miller et al.~\cite{Miller_2019,Miller_2021}, while the results by Riley at al.~\cite{Riley_2019,Riley_2021} give the upper limit $\Delta R_{(1.4-2.0)M_{\odot}} \simeq 0.32\,{\rm km}$. Reference~\cite{Raaijmakers_2021} investigated the additional effect on the EoS of the jointly estimated mass and radius of PSR J0740+6620 presented in~\cite{Riley_2021} by analyzing a combined data set from X-ray telescopes NICER and XMM-Newton. They concluded that $R_{1.4\,M_\odot}\sim R_{1.8\,M_\odot}\sim R_{2.0\,M_\odot}$ within 1\,km precision.

The lightest NS identified as the central compact object XMMU J173203.3-344518 could have a rather small radius $10.4^{+0.86}_{-0.78}$\,km within $1\sigma$ confidence according to Ref.~\cite{Doroshenko-light-HESS}. If confirmed, it would be an intriguing possibility of a superdense compact object different from a NS~\cite{Horvath2023}. However, as pointed in Ref.~\cite{Miao:2024vka-misc}, the results of Ref.~\cite{Doroshenko-light-HESS} would change, if the distance to the object is revised. The authors of Ref.~\cite{Miao:2024vka-misc} used the Gaia parallax measurements of the optical star and estimated the distance to the object to be shorter by factor 1.28. Consequently, they obtained the larger mass of $0.83^{+0.17}_{-0.13}\,M_\odot$ and the larger radius $11.25^{+0.53}_{-0.37}$\,km for XMMU J173203.3-344518.

\section{EoS and empirical constraints}

Typical hadronic EoSs are challenged by the requirement of simultaneous fulfillment of empirical constraints gained in studies of various nuclear systems: atomic nuclei, heavy-ion collisions, and astrophysics. Most difficult is to unite the description of the particle flow in heavy-ion collision requiring a soft EoS for the isospin-symmetric matter~\cite{Danielewicz:2002pu} and a large value of the maximum NS mass requiring a stiff EoS for the NS matter, see discussion in Ref.~\cite{Klahn-PRC74}. To resolve this problem in the framework of RMF models, the baryon-density dependence of hadron coupling constants was suggested in Refs.~\cite{Typel-Wolter-NPA656,Typel-PRC71}. In this case, the construction of thermodynamically consistent quantities requires additional care. In Ref.~\cite{KV-NPA759-KVOR} we proposed the RMF model with scaling of hadron masses and coupling constants, the SHMC model, in which hadron masses and meson–baryon coupling constants are dependent on the $\sigma$ mean field. For infinite nuclear matter, the scaling functions for masses and coupling constants enter the EoS only as a ratio, which dependence on the $\sigma$ field is chosen to gain the best description of empirical constraints that minimizes the number of fitted parameters. The $\sigma$-field scaling of RMF mass terms was motivated by experimental hints on the modification of hadronic masses and widths in hadronic matter and arguments for partial symmetry breaking with a baryon density increase. So, it looks natural that within the RMF approach not only baryon mass terms but also the mass terms of $\sigma$, $\omega$, $\rho$, and $\phi$ meson fields should be similarly dependent on the $\sigma$ field in the medium. In the SHMC models we deal with the usual Lagrangian approach and the derivation of thermodynamic quantities follows without ado. The models KVR and KVOR of such a type formulated in~\cite{KV-NPA759-KVOR} and named so in \cite{Klahn-PRC74}  allowed to satisfy most of the constraints on EoS known to that time from analysis of nuclei, heavy-ion collisions, and NSs, including the particle flow, DU, and maximum NS mass constraints, cf. Table V in Ref.~\cite{Klahn-PRC74}. To describe the new data on NS masses together with the flow constraint, two families of models (labeled as MKVOR and KVORcut) were constructed in Refs.~\cite{MKV-PLB748,MKV-NPA950-Hyp,KMV-NPA961-Delta} based on the KVOR model. The models implement various versions of the stiffening mechanism (the cut mechanism) developed in~\cite{MKV-PRC92-cut}, which aims at leveling off the sigma-field increase at some specified value. As a result, the effective baryon masses stop decreasing at some chosen value of the density $n_*\gsim (2-4) n_0$. Some microscopic support for such in-medium nucleon mass variation can be found in the renormalization group approach~\cite{Paeng-Lee-Rho-PRD88}. Also, if the decrease of the nucleon mass is determined by a decrease of the quark condensate in the medium, several mechanisms lead to a strong reduction of the condensate decrease rate at higher densities~\cite{Lutz-Friman-Appel-PLB474}. In the KVORcut family, the cut mechanism (strong variation of parameters with $\sigma$) is included in the $\omega$-field sector, whereas in the MKVOR family the cut-mechanism is implemented in the $\rho$-meson sector. Therefore, we could push up the maximum NS mass and simultaneously satisfy the particle flow constraint from heavy-ion collisions since there is no $\rho$ meson contribution for the isospin-symmetric matter.

The SHMC models were used in studies of heavy-ion collisions and NSs in Refs.~\cite{Khvorostukhin:2006ih,Khvorostukhin:2008xn,Khvorostukhin:2010aj,MKV-PRC92-cut,MKV-PLB748,MKV-NPA950-Hyp,KMV-NPA961-Delta,KMV-NPA970-rho,Grigorian:2017xqd,Grigorian:2018bvg,Maslov:2019dep}, demonstrating that the maximum NS mass is $M_{\rm max}>(2-2.2) M_{\odot}$ even in the presence of hyperons and $\Delta$ baryons. Also, the models fulfill the constraints on the EoS of the isospin symmetric matter from the nucleon flow and kaon production, giant monopole resonances, the constraints on the symmetry energy from neutron-proton elliptic flow difference measured by FOPI-LAND experiment, and nuclear analog isobaric states. The models describe appropriately optical nucleon potential $U(n)$ and for $n\lsim n_0$ we appropriately recover the results of the chiral perturbation theory. In the case of the beta-equilibrium matter the DU constraint is fulfilled. The NS cooling data are also properly described even with taking into account of hyperons~\cite{Grigorian:2017xqd,Grigorian:2018bvg}. The NS deformability calculated with our MKVOR-based models fits within the 90\% confidence region obtained from the GW170817 gravitational wave signal. For the KVORcut03-based model, the results lie on a border of the 90$\%$ confidence region~\cite{Maslov:2018ghi}. Thus, most of the presently known constraints are satisfied with the purely hadronic EoSs obtained with the SHMC model.

Usually, new degrees of freedom appearing in a phase transition lead to a decrease of a thermodynamic potential that necessarily results in the EoS softening. Nevertheless, the maximum NS mass remains above the modern empirical constraints in our SHMC models~\cite{MKV-PLB748,MKV-NPA950-Hyp,KMV-NPA961-Delta} even in the
presence of hyperons and $\Delta$'s. Reference~\cite{Maslov:2018ghi} also demonstrated the possibility that in the SHMC models the most massive NSs may contain a hadron-quark pasta phase and quark cores, satisfying the maximum NS mass constraint. In Ref.~\cite{Grigorian:2004jq} the hybrid star cooling scenario is shown to be compatible with the NS cooling data provided the density dependence of diquark gaps is taken into account and the nuclear medium cooling scenario is used for the description of the hadronic part of the NSs.

To avoid the EoS softening in the phase transition constructed by matching of thermodynamical potential, the authors of Ref.~\cite{Masuda_2013} suggested the so-called ``three-window scenario'' assuming an enforced transition from the purely hadronic phase to a crossover phase at some density $n_{\rm c1}\sim 2\,n_0$,
which at density $n_{\rm c2}\sim (4\mbox{--}7)\,n_0$ changes to the quark matter. By this logic one can use the nuclear equation of state only at densities $n< n_{\rm c1}$, some quark model at $n>n_{\rm c2}$, and a smooth interpolation in between. Such a picture could be supported by the quark percolation conjecture~\cite{Baym-PhysicaA96,Celik-Karsch-Satz-PLB97} assuming that quarks may begin to `jump' between nucleons at $n_{\rm c1}$ and, then, the fraction of `shared' quarks increases with a density increase. If were so, the hadron phase would change smoothly to the quarkish phase even though the latter one has a higher value of the thermodynamic potential (resulting in a stiffer EoS) than the hadronic one, see discussions in Refs.~\cite{Baym-TKPST-RPP81,Baym-Furusawa-ApJ885,Kojo-Baym-Hatsuda-ApJ934}.
Within the three-window scenario the hyperon and $\Delta$ puzzles get an almost trivial solution: as soon as the quark percolation starts, the formation of new baryonic states and the corresponding Fermi seas is forbidden. It should be noticed, however, that there is currently no quantitative description of the intermediate density phase. Moreover, the density interval $(2-4) n_0$ is well covered by low-energy heavy-ion collision experiments, which interpretation does not require the introduction of quark matter. Also, the NS masses obtained in ordinary RMF EoSs corresponding to central densities $(2-4) n_0$  are still low, $(0.7-1.5)M_{\odot}$.  Thus, within this approach, questions remain about the fulfillment of other constraints such as the flow constraint (covering densities $\lsim 4.5 n_0$) and the DU constraint suggesting  absence of the neutrino DU reactions in NSs with $M\lsim (1.35-1.5)M_\odot$, cf.~\cite{Klahn-PRC74}.

\section{Application of new NICER data}

Recently Ref.~\cite{Yamamoto:2023osc} proposed to interpret the NICER data in favor of approximate independence of the NS radii on the star mass in the interval of the NS masses between $1.4M_{\odot}$ and $2.0M_{\odot}$  and to use the relation
\begin{align}
R_{1.4M_\odot}\approx R_{2.0M_\odot}\,,
\label{R14-20-conjecture}
\end{align}
as a possible novel constraint on the NS EoS supported by the analysis of~\cite{Raaijmakers_2021}. A similar relation was discussed in Ref.~\cite{Han-Prakash-ApJ899}. From attempts to satisfy this constraint the authors of Ref.~\cite{Yamamoto:2023osc} concluded that it cannot be fulfilled with the purely hadronic EoSs softened by the admixture of hyperons, indicating thereby in favor of the existence of quark or hybrid phases in NS interiors.

\begin{figure}
\centering
\includegraphics[width=8cm]{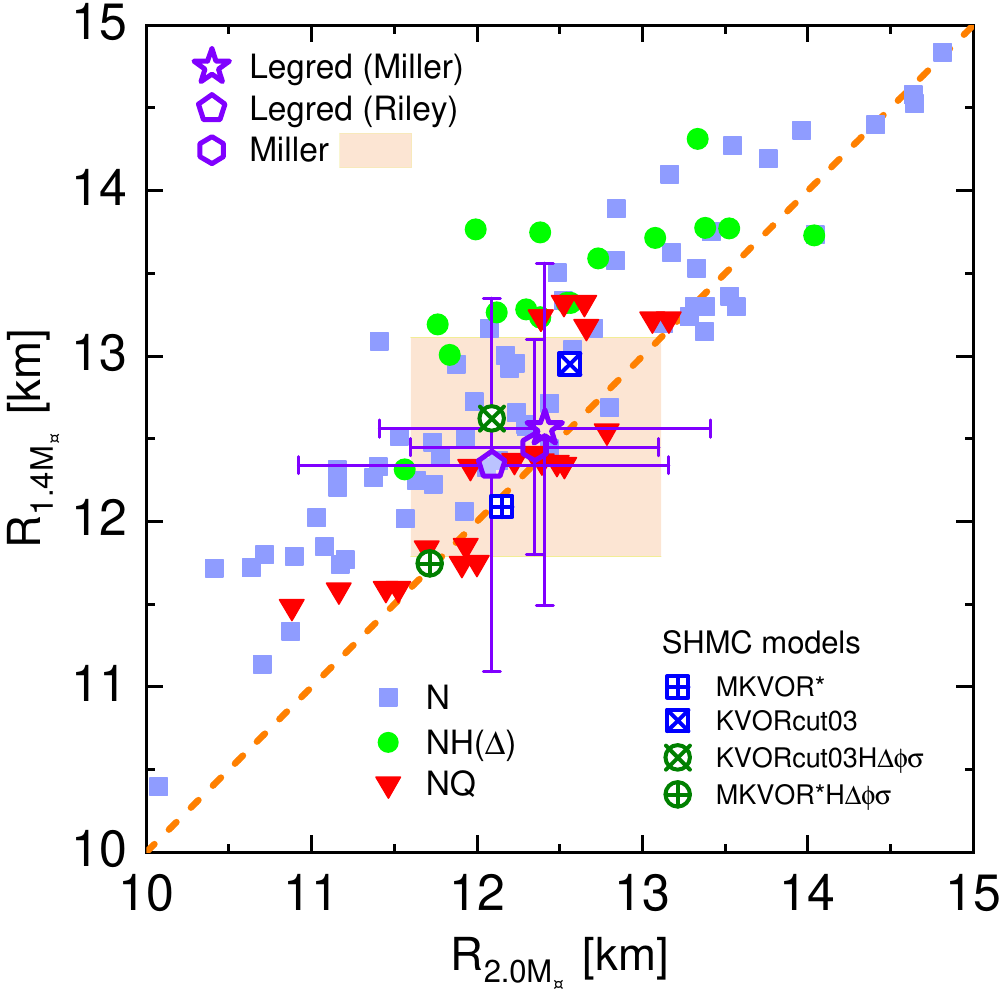}
\caption{Radii of NSs with masses 1.4\,$M_\odot$ and 2.0$M_\odot$ for the EoSs collected in Ref.~\cite{Ofengeim:2023nvc}.
Filled symbols stand for purely nucleonic EoSs (squares, N), EoSs with hyperons and/or $\Delta$ (circles, H($\Delta$)), and hybrid EoSs with nucleons and quarks (returned triangles, NQ). Symbols with error bars show the results by Legred et al.~\cite{Legred_2021}, Eqs.~(\ref{Legred-14},\ref{Legred-20}) using the NICER data \cite{Riley_2019,Riley_2021} by Riley et al.~(pentagon) and \cite{Miller_2019,Miller_2021} by Miller et al.~(star), and the results (\ref{Miller}) of the combined analysis from~\cite{Miller_2021}~(hexagon). The error bars of the latter analysis are visualized by the colored rectangle. The dashed line represents the relation $R_{1.4\,M_{\odot}}=R_{2.0\,M_{\odot}}$. The EoSs for the SHMC models are depicted by open symbols: squares are for purely nucleonic models -- MKVOR* (with the vertical cross) and KVORcut03 (with the diagonal cross); circles are for models with hyperons and  $\Delta$s -- MKVORH*$\Delta\phi\sigma$ (with the vertical cross) and KVORcut03$\Delta\phi\sigma$ (with the diagonal cross).
\label{fig:r14-20}}
\end{figure}

Before discussing the possibility of understanding new mass and radius measurements using SHMC  models of purely hadronic EoS, including both hyperons and $\Delta$s, we analyze typical EoSs for NSs used in the literature. We benefit from Ref.~\cite{Ofengeim:2023nvc}, where a set of cold NS EoSs was collected mainly from the CompOSE database~\cite{Compose} and from other works~\cite{Read-PRD79,Ozel:2016oaf,Ofengeim-PRD101}. All together, 162 EoSs were selected in~\cite{Ofengeim:2023nvc}. Dropping the EoSs with $M_{\rm max}<2.0\,M_\odot$ we remain with 103 EoSs. In Fig.~\ref{fig:r14-20} we show the radii $R_{1.4 \, M_{\odot}}$ and $R_{2.0\, M_{\odot}}$ calculated in the SHMC models and for the EoS collected in~\cite{Ofengeim:2023nvc}, among which there are 63 purely nucleonic EoSs (N), 18 EoSs with hyperons and/or $\Delta$s (NH($\Delta$)), and 22 hybrid EoSs with nucleons and quarks (NQ) depicted by squares, circles and triangles, respectively. Symbols with error bars show the results of the analyses~\cite{Legred_2021}, see Eq.~(\ref{Legred-20}), and~\cite{Miller_2021}, see Eq.~(\ref{Miller}). The colored rectangle visualizes the error bars given in (\ref{Miller}). The dashed line stands for Eq.~(\ref{R14-20-conjecture}).
We see that the radii for many purely nucleonic EoSs satisfy the constraint (\ref{R14-20-conjecture}), i.e. the corresponding squares lie close to the dashed line. Also, many N-EoSs produce radii $R_{1.4(2.0)\, M_\odot}$ falling within the large experimental error bars. Open squares with crosses show two versions of the purely nucleonic SHMC models. The minimal modification of the MKVOR model labeled MKVOR* in~\cite{KMV-NPA961-Delta} prevents the effective nucleon mass from vanishing at any density. The MKVOR* square is closer to the dashed line than that for KVORcut03, cf.~\cite{MKV-NPA950-Hyp}. In case of the MKVOR* model the radius difference $\Delta R_{(1.4-2.0)M_\odot}$ is negative, $-0.1$\,km, while for the KVORcut03 model it is positive $+0.4$\,km.  We stress that these two models differ in the density dependence of the symmetry energy, which is weaker in the first model for $n>n_0$.

The inclusion of hyperons (or/and $\Delta$s) shifts the radii (circles) in Fig.~\ref{fig:r14-20} up from the dashed line. For those H($\Delta$) EoSs whose radii agree with Legred's analyses, the radius difference $\Delta R_{(1.4-2.0)M_\odot}\gsim 1$\,km, and only one H($\Delta$) EoS is within error bars of the constraint~(\ref{Miller}), i.e. enters the colored rectangle. One more point satisfies this constraint marginally. Thus, indeed, among the EoSs collected in ~\cite{Ofengeim:2023nvc}, the inclusion of hyperons complicates the fulfillment of the condition (\ref{R14-20-conjecture}). On the contrary, many NQ EoSs satisfy well this condition, and many triangles lie close to the dashed line.
So, in favor of the statement~\cite{Yamamoto:2023osc} it is tempting to conclude that the quark admixture in the NS matter is necessary to reach the agreement with condition~(\ref{R14-20-conjecture}). However, the purely hadronic SHMC models with hyperons can also satisfy condition (\ref{R14-20-conjecture}). Open circles with crosses in Fig.~\ref{fig:r14-20} show the radii for the models MKVOR*H$\Delta\phi\sigma$ and KVORcut03H$\Delta\phi\sigma$~\cite{MKV-NPA950-Hyp,KMV-NPA961-Delta}. The suffix ``$\phi$'' means that in these models we included the $\phi$-meson mean field providing repulsion among hyperons and took into account the scaling of the mean-field $\phi$ meson mass term similar to the scaling of other mean-field mass terms and the nucleon mass. As shown in Ref.~\cite{MKV-PLB748} this scaling enhances the hyperon-hyperon repulsion and allows the solution of the hyperon puzzle. The suffix ``$\sigma$'' indicates that we include the effect of reducing hyperon-sigma coupling with a $\sigma$-field increase as it follows, e.g., from the quark-meson coupling model~\cite{Stone-QMC2007}. We see that the KVORcut03H$\Delta\phi\sigma$ point enters the colored rectangle in Fig.~\ref{fig:r14-20}, and MKVOR*H$\Delta\phi\sigma$ marginally satisfies this constraint. For the MKVOR*H$\Delta\phi\sigma$ model we have $\Delta R_{(1.4-2.0)M_\odot}=-0.03$ km and for KVORcut03H$\Delta\phi\sigma$ model, $0.5$\,km.

\begin{figure}
\centering
\includegraphics[width=8cm]{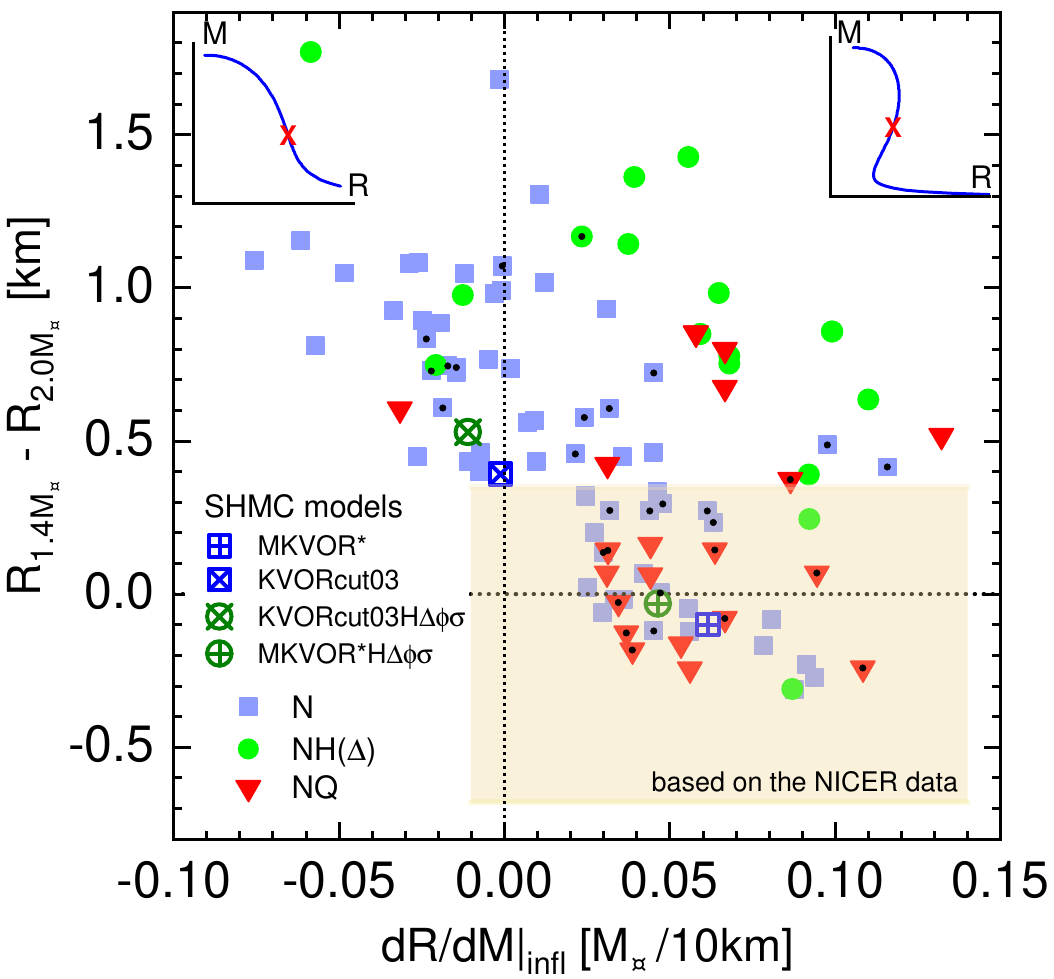}
\caption{Correlation between $\Delta R_{(1.4-2.0)M_{\odot}}$ and the form of the mass-radius curve characterized by the sign of the derivative $R'(M)$ at the inflection point (shown by crosses in the insertion plots as an illustration) for the same set of EoSs as in Fig.~\ref{fig:r14-20}. All symbols have the same meaning as in Fig.~\ref{fig:r14-20}. The rectangle indicates the interval $-0.68\,{\rm km}<\Delta R_{(1.4-2.0)M_{\odot}}<0.35$\,km encompassing the results of different analyses, see Eq.~(\ref{Legred-DR}) and the text below it. {Central dots indicate the EoSs, which fall within the colored rectangle in Fig.~\ref{fig:r14-20} satisfying the radius constraint from the combined analysis of Ref.~\cite{Miller_2021}.}
\label{fig:correlat}}
\end{figure}

The shape of the mass-radius curve can be characterized by the sign and magnitude of the derivative $R'(M)$ at the inflection point of the $R(M)$ curve, i.e. the point where $R''(M_{\rm infl})=0$. In Fig.~\ref{fig:correlat} we illustrate the correlation between $\Delta R_{(1.4-2.0)M_{\odot}}$ and $R'(M_{\rm infl})$ for different types of the EoSs.  If $R'(M_{\rm infl})<0$, then $M(R)$ is a monotonically decreasing function, shown in the left insertion plot in Fig.~\ref{fig:correlat}. If $R'(M_{\rm infl})>0$, then  $M(R)$ curve is non-monotonous with a backbending, as shown in the right insertion.  We see that most of the NH($\Delta$) and NQ EoSs give the $M(R)$ curve with backbending, while the nucleonic (N) EoSs can produce the $M(R)$ curves with and without backbending.
For MKVOR* and MKVOR*H$\Delta\phi\sigma$   EoSs, $R(M)$ curves have backbending and $R'(M_{\rm infl})>0$. These models are characterized by a softer density dependence of the symmetry energy than the KVORcut models. Respectively, the coefficient $L$ characterizing the density dependence of the symmetry energy nearby $n\simeq n_0$ is smaller ($L\simeq 41$ MeV) for MKVOR-type  models than that ($L\simeq 71$ MeV) for KVOR-type models.  The model  KVORcut03H$\Delta\phi\sigma$ demonstrates $R'(M_{\rm infl})<0$. The KVORcut03 EoS  shows a tiny negative value of $R'(M_{\rm infl})$. The  correlation between the density dependence of the symmetry energy and the shape of $M(R)$ curve is also visible in Fig.~1 of work~\cite{Kubis-PRC108-HESS} and in Fig.~5 of work~\cite{Hu:2020ujf} for the EoSs studied there.

The colored rectangle in Fig.~\ref{fig:correlat} indicates interval of $-0.68\,{\rm km}<\Delta R_{(1.4-2.0)M_{\odot}}<0.35\,{\rm km}$ motivated by the analyses \cite{Miller_2019,Miller_2021}. By central dotes we additionally mark those EoSs, whose NS radii occur within the colored rectangle in Fig.~\ref{fig:r14-20}. Only these EoSs agree with empirical data (\ref{Miller}) from the analysis~\cite{Miller_2021}. None of the NH($\Delta$) EoSs can be found simultaneously in the colored rectangles in Fig.~\ref{fig:r14-20} and in Fig.~\ref{fig:correlat}. The SHMC EoSs MKVOR* and MKVOR*H$\Delta\phi\sigma$   satisfy both constraints. The model KVORcut03 satisfies the constraint marginally. The model  KVORcut03H$\Delta\phi\sigma$ does not fulfill it.

\begin{figure}
\centering
\includegraphics[width=8cm]{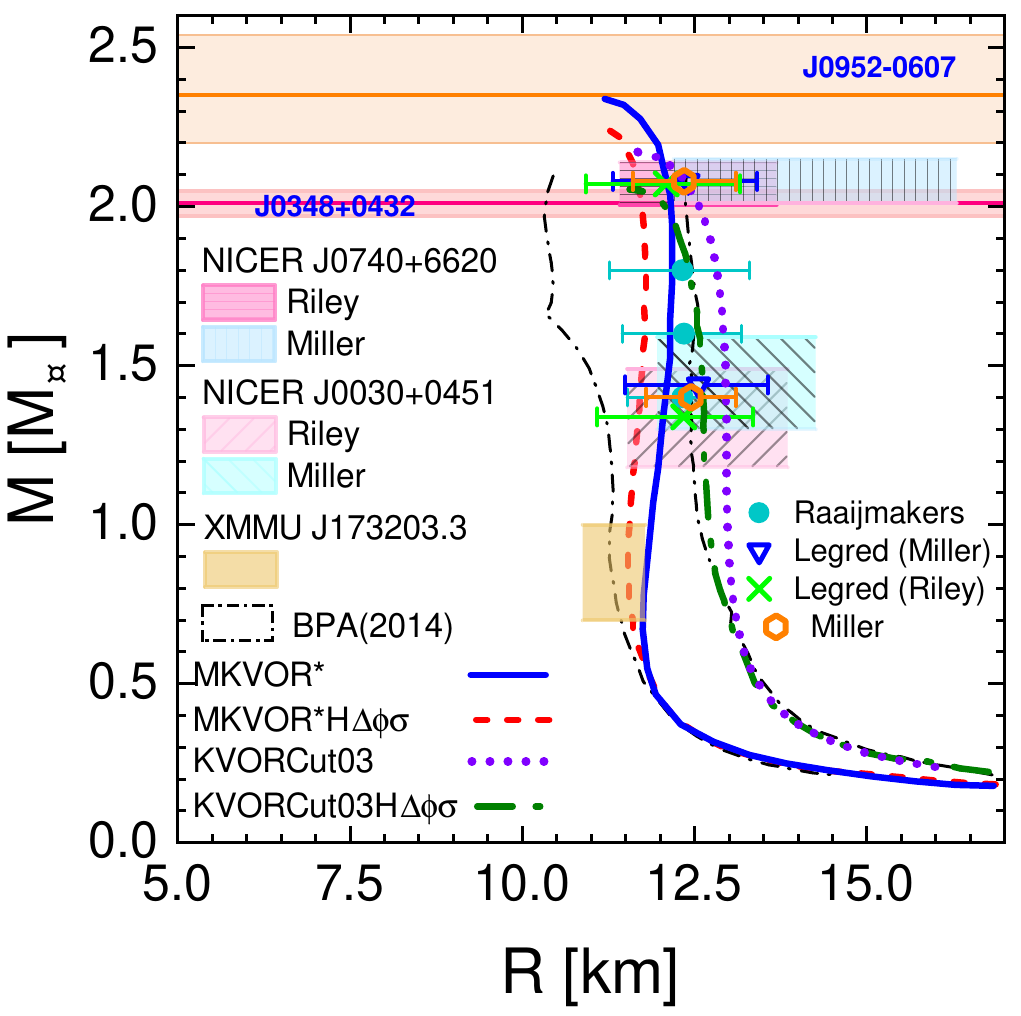}
\caption{ The mass-radius relation for cold non-rotating  NS. Solid, dashed, dash-dotted and dotted lines show the result for four SHMC models discussed in text.
Hatched regions show the results of NICER obtained by Miller et al. in~\cite{Miller_2019,Miller_2021} and Riley et al. in~\cite{Riley_2019,Riley_2021}, and the XMMU data after the distance correction in Ref.~\cite{Miao:2024vka-misc}. Dash-dotted contour shows the $M$--$R$ range from the Bayesian probability analysis (BPA)~\cite{Lattimer_2014}. Filled circles show the results of the analysis by Raaijmakers et al.~\cite{Raaijmakers_2021}. The results of the combined analysis of Legred et al.~\cite{Legred_2021} using data of Miller et al. and Riley et al.  are shown by triangle and cross symbols, respectively. Hexagons represent the analysis of the data done in Ref.~\cite{Miller_2021}.
\label{fig:nicer-radius}}
\end{figure}

Finally, let us compare the NS mass-radius relations for the considered SHMC models with the available empirical constraints, see Fig.~\ref{fig:nicer-radius}.
All four models have $M_{\rm max}$$>$$2.0\,M_\odot$. The NS mass constraint from the black widow pulsar J0952-0607 ($M_{\rm max}$$>$$2.2\,M_\odot$) is satisfied by the purely nucleonic MKVOR* model and also by the MKVOR*H$\Delta\phi\sigma$ model with hyperons and $\Delta$s. With account for the rotation correction~\cite{Brandes:2023hma} the lower edge of the constraint can be reduced to $2.1\,M_\odot$, then also the KVORcut03 EoS and marginally the KVORcut03H$\Delta\phi\sigma$ EoS will satisfy it. We stress that the mass-radius curves for all four SHMC EoSs go through the results of extended analyses of the NICER data~\cite{Legred_2021,Raaijmakers_2021} including additional empirical and theoretical information.
As for the results of the direct NICER measurements, the SHMC EoSs do not pass through the radius range obtained by Miller et al.~\cite{Miller_2021} for the $2.0\,M_\odot$ NS, however all agree with the range deduced by Riley et al.~\cite{Riley_2021}.
For the direct measurements presented in Refs.~\cite{Riley_2019,Miller_2019}, the MKVOR*H$\Delta\phi\sigma$ model does not pass through the 1$\sigma$ range suggested for $R_{1.4\,M_\odot}$ in Ref.~\cite{Miller_2019}, but agrees with the result~\cite{Riley_2019} and almost touches the $1\sigma$ error bar (hexagon in Fig.~\ref{fig:nicer-radius}) given in the analysis of~\cite{Miller_2021}, see Eq.~(\ref{Miller}). We note also that the re-analysis in Ref.~\cite{Miller_2021}, Eq.~(\ref{Miller}), reduces the tension between the direct NICER results by Miller et al.~\cite{Miller_2019}, $R_{1.4\,M_\odot}=13.02_{-1.06}^{+1.24}$\,km, and the constraint put on $R_{1.4\,M_\odot}$ in Ref.~\cite{Capano-Tews-NatAstr4}, where the authors used the chiral effective field theory and the gravitational wave observations of the binary NS merger GW170817. They claim $R_{1.4M_{\odot}}=11.0^{+0.9}_{-0.6}$ km with 90\% confidence. The XMMU rectangle shows the constraint obtained after the distance correction in Ref.~\cite{Miao:2024vka-misc}. The MKVOR-based models satisfy this constraint unlike the KVORcut-based models. This difference may be related to the weaker density dependence of the symmetry energy in the former ones. In Fig.~\ref{fig:nicer-radius} we also see that stars with hyperons and $\Delta$s have smaller radii than stars without.

\section{Conclusion}

At the hand of a set of 103 EoSs collected in Ref.~\cite{Ofengeim:2023nvc} from the CompOSE database~\cite{Compose} and from Refs.~\cite{Read-PRD79,Ozel:2016oaf,Ofengeim-PRD101}, among which there are 18 EoSs with hyperons and/or $\Delta$s and 22 hybrid NQ EoSs, we demonstrated that indeed most of the used hadronic EoSs with hyperons and/or $\Delta$s (NH($\Delta$)) do not satisfy the condition conjectured in~\cite{Yamamoto:2023osc} that the stellar radius  weakly changes with the mass increase from 1.4 to $2.0\,M_\odot$, Eq.~(\ref{R14-20-conjecture}).
According to the analyses of a combined data set from the X-ray telescopes NICER and XMM-Newton supplemented by the gravitational wave constraints and theoretical constraints on the EoSs~\cite{Legred_2021,Miller_2021,Raaijmakers_2021} discussed above, the radius difference is limited as $-0.68\,{\rm km}<\Delta R_{(1.4-2.0)M_{\odot}}<0.35$\,km. None of the NH($\Delta$) EoS from the EoSs collection in~\cite{Ofengeim:2023nvc} satisfies both this constraint
and the constraint~(\ref{Miller}) on stellar radii obtained in the analysis~\cite{Miller_2021}. On the contrary, the hybrid EoSs with nucleons and quarks constructed within the three-window scenario fulfill these constraints, see Fig.~\ref{fig:r14-20}.

We analyzed the shape of the mass-radius curve, $M(R)$, for different EoSs and demonstrated that for the NH($\Delta$) and NQ EoSs it has, as a rule,  backbending, as shown in Fig.~\ref{fig:correlat}. We also argued that this shape  favors  a smoother density dependence of the nuclear symmetry energy and a smaller value of $L$.

We showed that the hadronic relativistic mean-field models with the $\sigma$-field-scaled hadron masses and coupling constants (SHMC), --- the KVORcut03- and MKVOR-based models, --- constructed in Refs.~\cite{MKV-PLB748,MKV-NPA950-Hyp,KMV-NPA961-Delta} pass most of the currently known constraints from experiments with nuclei, heavy-ion collisions and compact stars, including the analyses of the new NICER mass-radius measurements. The MKVOR-type models with and without hyperons and $\Delta$s satisfy well the condition~(\ref{R14-20-conjecture}) conjectured in~\cite{Yamamoto:2023osc}. The KVORcut03 EoS satisfies it marginally. The mass-radius curves for the considered SHMC models shown in Fig.~\ref{fig:nicer-radius} agree very well with the constraints based on the new NICER data.

A weak variation of the NS radius of the star mass for $M>0.5 M_{\odot}$ has been noticed already in works~\cite{MKV-PLB748,MKV-NPA950-Hyp,KMV-NPA961-Delta} and it is now supported by the NICER data and suggested as a constraint in Ref.~\cite{Yamamoto:2023osc}. We hope that our results can be treated in favor of RMF models including $\sigma$ scaling not only of baryon masses but also of meson masses.

The presented analysis shows that the new NICER-data-based constraints on the NS radii are very selective to the EoS used in the NS descriptions. The future next-generation gravitational-wave observatory~\cite{Evans:2021gyd} -- Cosmic Explorer -- is planned to detect NS mergers with a high rate that will enable the determination of stellar radii with a very high precision of 100 meters. Thus, the conjecture of Ref.~\cite{Yamamoto:2023osc} supported by our results \cite{MKV-PLB748,MKV-NPA950-Hyp,KMV-NPA961-Delta} obtained with purely hadronic SHMC models and other results illustrated in Figs.~\ref{fig:r14-20}--\ref{fig:nicer-radius} could be, hopefully, verified experimentally.

\acknowledgments

We are grateful to Dmitri Ofengeim for discussing and sharing with us his collection of EoSs and mass-radius curves from Ref.~\cite{Ofengeim:2023nvc}. The analysis of hadronic EoSs is carried out within the framework of the Russian Science Foundation program under grant RSF~21-12-00061. The work is also partially supported by the grant VEGA~1/0353/22.

\bibliography{refs-NICER}

\end{document}